\documentclass[aps,prd,onecolumn,groupedaddress,showpacs,nofootinbib,amssymb]{revtex4}
\usepackage{graphicx,bm}
\usepackage[english]{babel}
\usepackage{amsmath}
\usepackage{amssymb}
\usepackage{amsfonts}

\begin{document}

\def\cL{{\cal L}}
\def\be{\begin{eqnarray}}
\def\ee{\end{eqnarray}}
\def\bea{\begin{eqnarray}}
\def\eea{\end{eqnarray}}
\def\beq{\begin{eqnarray}}
\def\eeq{\end{eqnarray}}
\def\tr{{\rm tr}\, }
\def\nn{\nonumber\\}
\def\e{{\rm e}}


\title{A non-polynomial gravity formulation for Loop Quantum Cosmology  bounce    }

\author{Stefano Chinaglia $^{1}$,
\footnote{E-mail address: s.chinaglia@unitn.i },
  Aimeric Colléaux  $^{1}$,
\footnote{E-mail address:aimeric.colleaux@unitn.it  }
and Sergio~Zerbini$^{1}$
\footnote{E-mail address: sergio.zerbini@unitn.it}
}
\affiliation{
$^1$ Dipartimento di Fisica, Universit\`{a} di Trento, Italy, and TIFPA, Istituto Nazionale di Fisica Nucleare, Trento, Italy}


\begin{abstract}

Recently the so-called mimetic gravity approach has been used to obtain  corrections to Friedmann equation of General Relativity similar to the ones present in loop quantum cosmology. In this paper, we propose  an alternative way to derive this modified Friedmann equation via the so-called non-polynomial gravity approach, which consists in adding geometric non-polynomial higher derivative terms to Hilbert-Einstein action, which are nonetheless polynomials and lead to second order differential equation in Friedmann-Lema\^itre-Robertson-Walker spacetimes. Our explicit action turns out to be  a realization of the Helling proposal of effective action with infinite number of terms. The model is investigated also in presence of non vanishing cosmological constant and a new exact bounce solution is found and studied.
  
\end{abstract}


\maketitle


\section{Introduction}

It is believed that singularities are not part of nature because they indicate a breakdown of predictivity of the theory under consideration. It is also well known that General Relativity (GR) plus ordinary matter admits solutions for the space-time metric
which are singular. For our purposes, a space-time metric  will be singular if there exists ill defined curvature invariants
at some points. Simple and familiar examples are the static  Schwarzschild metric and the GR solution with ordinary matter or radiation in a Friedmann-Lema\^itre-Robertson-Walker (FLRW) space-time. Of course, this is not the most general definition of singular space-time (see for example \cite{Li} and references therein), but it will be sufficient for us.

In this paper, we will be only interested in working within flat FLRW space-times. One possible approach to face this singularity issue may consist in considering high energy corrections to GR action in order to cure this divergence. Different attempts have been proposed to deal with these divergences, essentially in three directions: either taking into account  quantum gravity corrections, or considering  new kinds of matter, or by modifying GR (what is usually equivalent to the introduction of new fields). As a consequence, the resulting modified Friedmann equations may contain regular solutions, for example bounces.  The number of the papers dealing with cosmological bounces is quite huge. A partial list of papers is \cite{Tu1,Tsu,add,No,Bra,Ea,od,B,lo,staro,Gie,I}. Some recent progresses in the general features of the problem have also been presented in \cite{Li,qiu,Dz}, \cite{Malkiewicz}, \cite{Malkiewicz_2},\cite{NO17} and \cite{Cai}.

 Within the specific approaches of string theory and loop quantum cosmology (LQC),  models with no singularities in their cosmological sector have been proposed: see e.g. \cite{bo}, \cite{Ashtekar}, \cite{Ashtekar_2} for LQC and \cite{Helling}, \cite{Helling_2} for string theory.

With regard to the second proposal,  Helling \cite{Helling_3} and independently Date and Sengupta \cite{Date}, suggested a modification of GR Lagrangian which gives the same correction to the Friedmann equation than LQC, with therefore the same  bounce. This approach was intended to be an effective action formulation of the loop quantization procedure of FLRW space-times. Helling showed that a formulation in terms of an infinite sum of curvature invariants is possible, but it was not possible to write it explicitly.

 More recently, Chamseddine and Mukhanov working within the so called "mimetic approach" \cite{mi_1, Muk0} (see also the similar construction in \cite{V}) followed this idea, and  in two papers \cite{Mukhanov,Mukhanov_2}  made use of a non-polynomial function of the mimetic field in a simple manner, and were able to reproduce the LQC result. Note that within the mimetic approach, but including in the action a suitable potential  for the mimetic field, it is possible to show the existence of cosmological bounces. Examples are provided in \cite{Muk0,mi_6}. Furthermore, the bounce mimetic approach has recently been generalized in \cite{H}. Other recent papers on bounce loop cosmology are \cite{Z,Z1}, while mimetic modified gravity is discussed in \cite{Odi15}.  

 Furthermore, it should be stressed that dealing with modified gravity models on FLRW space-times, in general other singularities may arise (see for example \cite{NOT} and references therein). 

 In this paper, we propose to implement the Helling construction by finding an explicit Lagrangian built only from the metric field  and that leads to the LQC corrections. This Lagrangian is constructed via the so-called non-polynomial gravities (NPG)\cite{Deser}. The NPG approach  is intended to mimic a specific sector of a fundamental (i.e. background independent) effective theory, in which only  gravitational metric corrections with no additional derivatives are present. In this way,  invariants built making use of  non-polynomial terms in the metric  become  polynomials in the FLRW sector, becoming  candidates to build an effective action there.

The paper is organized as follow. In Section 2,  after expliciting the construction of the scalars, the action is written  and the
 associated equations of motion are derived.  In Section 3,  in presence of  a cosmological constant and a perfect fluid,  the exact solutions of the model are found and discussed. The paper ends with Section 4, in which a discussion on our results is presented.

\section{Action \& equations of motion}

To begin with, let us consider a flat Friedmann-Lema\^{i}tre-Robertson-Walker metric (FLRW) $\bar{g}$ defined by the following space-time interval :  
\begin{eqnarray}
\label{FLRW}
 ds^2 = - N(t) dt^2 + a(t)^2 d\vec{x}^{\,2}.
\end{eqnarray}
Here $N(t)$ is an arbitrary function which implements the time reparametrization invariance.

We want to build an effective action that reproduces some quantum geometry corrections. For this reason, we will be interested by scalars that are built from a particular geometric property of FLRW space-times, namely that the following projector : 
\begin{eqnarray}
\tau^{\alpha}_{\beta} =\delta^{\alpha}_{t} \delta^{t}_{\beta}  = \text{diag}( 1,0,0,0),
\end{eqnarray} 
is actually a true tensor, and that the quantity $\sqrt{N} \delta_\alpha^{t} $ is a true vector in FLRW. In order to see why, we can provide explicit tensorial forms to these objets. Consider the following vector and tensor : 
\begin{eqnarray}
\label{Order0}
  V_\alpha := \frac{\partial_{\alpha} R}{\sqrt{-\partial_{\sigma} R\partial^{\sigma} R} }   \;\;\;\;\; \text{and} \;\;\;\;\;   V_{\alpha\beta}  :=  V_{\alpha} V_{\beta}  \, .
\end{eqnarray}
For the considered metric, these geometric tensors are of order-$0$, namely, they do not depend on the derivatives of the metric (here the scale factor). Indeed denoting the restrictions of the tensors (\ref{Order0}) on (\ref{FLRW}) by $ V \bigr\rfloor = V  \bigr\rvert_{g=\bar{g}} \, $, etc..., one can see that they are indeed order-$0$ tensors with the claimed geometrical interpretation :
\begin{eqnarray}
  V_\alpha  \bigr\rfloor = -\sqrt{N} \delta_\alpha^{t}     \;\;\;\;\; \text{and} \;\;\;\;\;   V_{\alpha\beta}  \bigr\rfloor = -\tau_{\alpha\beta} 
  \end{eqnarray} 
This property follows from  the fact that for any scalar $Q$, $\partial_{\alpha} Q$ has only one component when evaluated on (\ref{FLRW}). We have chosen $Q=R$ here for simplicity. It is exactly the same type of property that the Weyl and Cotton tensors have in spherical symmetry \cite{Deser, Zerb, ZerbAim}, except that here the property is quite trivial. See \cite{RBHNPG,Deser2} for more details in the case of spherical symmetry.

In all  classes of space-times that share this property, like spherical symmetric space-times or Bianchi type I, one can build scalars from these tensors  that will be second order in these classes, in particular in (\ref{FLRW}), but higher order otherwise.  The two second order invariants we shall be interested in are 
\begin{eqnarray}
\label{Scalars}
 K  := \frac{1}{9} \big( \nabla^{\alpha}\nabla^{\beta} V_{\alpha\beta} - V_{\alpha} \nabla^{\alpha} \nabla^{\beta} V_{\beta} \big) \;\;\;\;\; \text{and} \;\;\;\;\;  \Omega  := \frac{R}{6}-2 K \,.
\end{eqnarray}
With these invariants we may construct gravity models denoted NPG in \cite{Deser}, and so we will use this name here. Furthermore, these scalars are chosen so that their restrictions to (\ref{FLRW}) are :
\begin{eqnarray}
  K \Bigr\rfloor =\frac{H^2}{N}  \;\;\;\;\; \text{and} \;\;\;\;\;  \Omega \Bigr\rfloor  =\frac{\dot{H}}{N} -  \frac{H \,
    \dot{N}}{2 N^2} \,,
\end{eqnarray}
because, together with the Ricci scalar, they form a basis of order-2 scalars in FLRW space-times, and setting $N(t)=1$, they are actually the simpliest ones.
Here, $H$ is the Hubble parameter, and $\dot{H}=\frac{d H}{dt}  $. Note that working in flat FLRW space-times, there exists other invariants, which have similar properties, see for example \cite{gao, ZerbAim}, but the ones we have chosen are also relevant in spherically symmetric space-times \cite{RBHNPG}.

We recall  that, in principle, it is possible to reproduce the loop quantum cosmology modification of Friedmann equation, and therefore the bounce that replaces the big bang, via higher order corrections to Einstein-Hilbert action \cite{Helling_3,Date}. These corrections have to lead to second order equations of motion, as shown by Helling, and so are truly geometrical corrections, in the sense that, unlike a generic modified gravity model, they do not involve additional fields with no direct geometrical meaning compared to the metric, or, for example, compared to the scalar field responsible to the local rescaling invariance in some models of conformal gravity.

In the  paper \cite{Helling_3}, it was also shown that such corrections are possible to write as an infinite series of polynomials of contractions of Ricci tensors, even though it was not possible to write this effective action explicitly. In our approach, making use of the two scalars (\ref{Scalars}) defined above, a possible way to achieve this task is to start with  the following action : 

\begin{eqnarray}
\label{Action}
 \mathcal{I} =\int d^4x \sqrt{-g} \left( \frac{R -2 \Lambda + \Big[ L_{NPG}^{\infty}\Big] }{2\kappa}+L_m \right),
\end{eqnarray}
where $\kappa=8\pi$, with the Newton constant $G=1$,  $\Lambda$ is the cosmological constant, $L_m$ is the Lagrangian density of matter, and
\begin{eqnarray}
\label{NPG}
L_{NPG}^{\infty}= -2 \Omega + \frac{4 \Omega}{ S} \Bigg( 1- \sqrt{1- S}  \, \Bigg)\,.
\end{eqnarray}
Here we have introduced the dimensionless scalar $S=\frac{3}{2 \pi \rho_c}\, K$, with $\rho_c$ playing the role of critical density, which in our approach is a {\it free} dimensional parameter. 

Some comments are in order. This contribution, which modifies the GR term, is similar to Born-Infeld Lagrangian for non linear electrodynamics and does not follow from first principles. On the other hand, it may be interpreted  as
\begin{eqnarray}
\label{NPG1}
L_{NPG}^{\infty}=-4 \sum_{i=0}^{\infty } (-1)^{i+1}  \binom{1/2}{1+i} \; S^i \; \Omega\,,
\end{eqnarray}
where $\binom{n}{m}$ is the generalized binomial coefficient defined by $\binom{n}{m} :=\frac{\Gamma (n+1)}{\Gamma  (m+1) \Gamma  (n-m+1)}$. The bracket in equation (\ref{Action}) are used in order to emphasize that, within this kind of minisuperspace approach, one can only hope to find the desired scalar up to scalars that vanish or are boundary terms (at least) in the class of space-times in which the reconstruction is done, in our case in flat FLRW (\ref{FLRW}). For example, one could add scalars involving the Weyl tensor or background dependent boundary terms (as those of \cite{ZerbAim}) in the action without modifying the dynamics of (\ref{Action}) for FLRW space-time.   Therefore, $L_{NPG}^{\infty}$ is only a particular NPG representative of an infinite class of scalars (that includes polynomial ones like in \cite{Helling_3}) with equivalent contributions to the equations of motion in (\ref{FLRW}).

Furthermore, for this specific space-time, the additional term, despite its non-polynomiality, may be considered in (\ref{NPG1}) as an infinite sum   of  polynomials in the metric, and therefore (\ref{Action}) constitutes  a suitable effective action, whose coupling constants are fixed in order to reproduce the LQC modification of Friedmann equations. Note also that the $i=0$  term of the sum (\ref{NPG1}), namely $2 \Omega$,  is equivalent to the Ricci scalar in FLRW, since they differ from each other by a total derivative. Moreover, in its present form, the correction $L_{NPG}^{\infty}$ seems of higher order,
but it is in fact equivalent, up to (background dependent) boundary terms, to the correction of \cite{Helling_3,Date}. Indeed, they differ from each other by a total derivative :
\begin{eqnarray}
\Bigg( R -2 \Omega + \frac{4 \, \Omega}{S} \Big( 1- \sqrt{1- S} \Big) \Bigg) \Biggr\rfloor   =  8 \pi \rho_c \Bigg( 1- \sqrt{1-S} - \sqrt{S} \arcsin(\sqrt{S}) \Bigg) \Biggr\rfloor+ \frac{4}{\sqrt{-g}} \sqrt{\frac{2 \pi \rho_c}{3}}  \dot{B}   \Biggr\rfloor ,
\end{eqnarray} 
with
 \begin{eqnarray}
B= \frac{\sqrt{-g}}{\sqrt{N}} \left(\csc^{-1}\left(\frac{1}{\sqrt{S}}\right)-\frac{1-S-\sqrt{1-S}}{\sqrt{S}}\right).
\end{eqnarray}
Note that in both cases, the GR contribution is cancelled, because $\sqrt{-g}\big(R-2\Omega\big) \bigr\rfloor = \frac{d}{dt}\Big( \frac{4 \, a^3 H}{\sqrt{N}} \Big) $, and what is left is only a non-polynomial effective action and an effective cosmological constant $8 \pi \rho_c$  in the first order form of the right-hand-side. Therefore, in FLRW and up to boundary terms, the sequence (\ref{NPG1}) of polynomial curvature scalars is the only one that gives the LQC modification of Friedmann equation, as we will see now. 
\\
\\
Making use of a minisuperspace approach (Weyl method), from  ansatz (\ref{FLRW}) and action (\ref{Action}) we can derive the Euler-Lagrange (EL) equations of motion by making the variation with respect to Lagrangian coordinates  $N(t)$ and $a(t)$ . The Principle of Symmetric Criticality applied to the isometry group of an homogeneous and isotropic universe assures that the reverse process (the right one) will give the same results \cite{PSC1,PSC2}. 

We also assume that the matter is a perfect fluid, with equation of state $p=w \rho$, $\rho$ and $p$ being the density and the pressure. Making the variation with respect to $N(t)$, one gets the Friedmann equation, and by setting $N(t)=1$ after the variation, one has
\begin{eqnarray}
\label{EOM}
4 \pi  \rho_c \left(1-\sqrt{1-\frac{3 H^2 }{2 \pi  \rho_c}}\right) =8 \pi   \rho+\Lambda\, .
   \end{eqnarray}
As a first check, when $\frac{H^2}{\rho_c} \ll 1$, one recovers the Friedmann equation of GR.

Defining $\bar{\rho} := \frac{ \Lambda}{8\pi} + \rho $, one gets the standard form of the LQC corrected Friedmann equation : 
 \begin{eqnarray}
H^2=\, \frac{8 \pi \bar{\rho}  }{3} \Big( 1 -  \, \frac{ \bar{\rho}  }{\rho_c} \Big).
\end{eqnarray}
 Making the variation with respect to $a(t)$, one gets the other Friedmann equation, which contains the acceleration. For our purposes, we do not need it since it can derived from (\ref{EOM}) and the energy conservation equation
 \begin{eqnarray}
 \label{Cons}
\frac{d\rho}{dt}+3H \big(\rho + p \big)=0\,,
\end{eqnarray}
consequence of the diffeomorphism invariance of our invariant action. Thus, only two of these three equations are independent, and  one may use only   equations  (\ref{EOM}) and  (\ref{Cons}).

  \section{Exact solutions for general equation of state parameter  $w$ and cosmological constant $\Lambda$}
    
  We recall the  equation of state for the perfect fluid $p=w \rho$. Then, introducing for the sake of simplicity $ \tilde{\rho}=8 \pi  \rho$ and $\mu=\frac{1}{8 \pi  \rho_c}$, one has
 \begin{eqnarray}
 \begin{split} 
 \label{SET}
3 H^2 &= (\tilde{\rho} + \Lambda )- \mu ( \tilde{\rho} + \Lambda)^2,\\
 \frac{d \tilde{\rho}}{\tilde{\rho} } &=-3 (1+w) H dt.
\end{split} 
\end{eqnarray}
 First we note that, without solving the differential equation, it is possible to show that a bounce solution is present, namely there exists $a_* >0$ such that $H_*=0$ and $\dot{H_*}>0$. In fact the first equation on the bounce $H_*=0$ gives the condition $1-\mu \Lambda=\mu\tilde{\rho_*} $, namely $\mu \Lambda<1$, which is therefore a necessary condition.

 We now derive the exact solution.  Inserting the first equation into the second one leads to :
 \begin{eqnarray}
\frac{d X} {(X-\Lambda) \sqrt{ X -\mu X^2}}= \pm \sqrt{3} (1+w)dt\,.
\end{eqnarray} 
where $X=\rho + \Lambda$. Thus, 
\begin{eqnarray}
\frac{2 \tanh ^{-1}\left(\frac{\sqrt{\Lambda } \sqrt{1-X
   \mu }}{\sqrt{X} \sqrt{1-\Lambda  \mu
   }}\right)}{\sqrt{\Lambda } \sqrt{1-\Lambda  \mu }}=\pm \sqrt{3}  (1+w) t+c \, , 
\end{eqnarray}
where $c$ is the integration constant. In the following, we may put $c=0$ without any problem. Solving in $X$ and thus in $\tilde{\rho}$ gives :  
\begin{eqnarray}
\tilde{\rho}(t)=-\frac{2 \Lambda  (-1+\Lambda  \mu )}{-1+2 \Lambda  \mu
   +\cosh \left(\left(\pm \sqrt{3} t (1+w)  \right)
   \sqrt{\Lambda } \sqrt{1-\Lambda  \mu }\right)}.
\end{eqnarray}
The second equation of (\ref{SET}) admits the usual well known solution  $a=a_0 \tilde{\rho}^{\frac{-1}{3(1+ w)}}$. As a consequence, one has
 
\begin{eqnarray}
\label{SOLUTION}
a(t) =a_0 \left(\frac{-1+2
   \Lambda  \mu +\cosh \Big(\left(\sqrt{3}  (1+w) t
   \right) \sqrt{\Lambda } \sqrt{1-\Lambda  \mu
   }\Big)}{2 \Lambda  (1-\Lambda  \mu )}\right)^{\frac{1}{3 (1+w)}}.
\end{eqnarray}
Here, we recover the condition $1-\mu \Lambda>0$,  $\Lambda >0$. Given this solution, one can check that the scalar $\partial_\sigma R \partial^\sigma R$ is not vanishing everywhere, and the scalars (\ref{Scalars}) are indeed well defined. 

As a further check of the solution, we can study the two limits $\mu \to 0$ and $\Lambda \to 0$. First, the GR limit, namely
\begin{eqnarray}
\begin{split}
& \underset{\mu \to 0}{\lim} \;  \tilde{\rho}(t) = \Lambda \;  \text{csch}^2\Big( \frac{1}{2} \sqrt{\Lambda }( \sqrt{3} t ( 1 + w)  \Big),
\\
& \underset{\mu \to 0}{\lim} \; a(t) = a_0 \Big( \frac{1- \cosh(\sqrt{\Lambda }( \sqrt{3} t ( 1 + w) )}{2 \Lambda} \Big)^{ \frac{1}{3(1+w)}},
\end{split}
\end{eqnarray}
This is the solution of GR with non vanishing cosmological constant, and one recovers the Big Bang solution at $t=0$.

In the other limit, one has 
\begin{eqnarray}
\begin{split}
& \underset{\Lambda \to 0}{\lim} \;  \tilde{\rho}(t) = \frac{4}{ \big( \sqrt{3} t (1+w) \big)^2 + 4 \mu},
\\
& \underset{\Lambda \to 0}{\lim} \; a(t) =a_0 \Big(\mu +\frac{1}{4} ( \sqrt{3}t(1+w) )   \Big)^{\frac{1}{3(1+w)}}.
\end{split}
\end{eqnarray}
and one recovers the original LQG bounce solution in absence of cosmological constant.

Now we study our exact solution with respect to the coordinate time $t$. We already have shown the existence of the bounce. In particular,  for $t$ small, one has,    
\begin{equation}
\label{limit_t_2}
a(t \rightarrow 0) = a_0 \left( \frac{\mu}{1 - \mu \Lambda} \right)^{1/3(1+w)} \left( 1 + \frac{(1 - \mu \Lambda) (1 + w)}{4 \mu} t^2 + ... \right) \ .
\end{equation}
We see that the  minimal value is $a(0) = a_0 \left( \frac{\mu}{1 - \mu \Lambda} \right)^{1/3(1+w)}$, corresponding to the bounce. Moreover, already eq. (\ref{SOLUTION}) shows that there $a(t)$ is never vanishing: indeed, the hyperbolic cosine is always greater than 1, so $\cosh{x} - 1 \geq 0$; and since $\mu$ and $\Lambda$ are both positive, the scale factor is always positive and never vanishing.

The other interesting limit is the one for $t$ very large. Since we already have taken  the cosmological
constant into account, we take $w >-1$. We remind that $\cosh{x} \rightarrow e^{|x|}$, for $x \rightarrow \pm \infty$, and one has
\begin{equation}
\label{limit_t_3}
a(t \rightarrow \infty) = \frac{a_0}{(2 \Lambda (1 - \mu \Lambda))^{1/3(1+w)}} \left( 2 \mu \Lambda - 1 + \exp{ \left( \sqrt{3 \Lambda (1 - \mu \Lambda)} (1+w) t \right) }\right)^{1/3(1+w)},
\end{equation}
 the exponential becomes dominant corresponding to an accelerating universe. Thus, our solution may represent dark energy (DE), with a chosen suitable scale, and for large $t$ \cite{Reiss}, \cite{Perlmutter}.

 We conclude this Section discussing the limits $\mu$ and $\Lambda$  large. We have seen that the product $\mu \Lambda$ must be $\mu \Lambda < 1$. This is not a problem for DE issue because $\mu=\frac{1}{8\pi \rho_c}$ mimics a quantum correction and thus it can be taken small, because $\rho_c$ is very large, and for DE $\Lambda$ is small. The situation is different with $\Lambda$ not small, as during inflation, and our solution, in this case, may not be interesting.

Finally,  concerning the scalars used in the construction, given their non-polynomial forms, one could wonder if they are regular at the bounce, like polynomial scalars. One can check that given the solution (\ref{SOLUTION}), their behaviours are :  
\begin{eqnarray}
\begin{split}
&\underset{t \to 0}{\lim} \;   \partial_\sigma R \partial^\sigma R = \underset{t \to 0}{\lim} \;  K  = 0,
\\
&\underset{t \to 0}{\lim} \;   \nabla^{\alpha}\nabla^{\beta} V_{\alpha\beta} =\underset{t \to 0}{\lim} \;  V_{\alpha} \nabla^{\alpha} \nabla^{\beta} V_{\beta} =  3 \, \underset{t \to 0}{\lim} \; \Omega= \frac{-3 (1+w) (-1 + \Lambda \mu) }{2 \mu},
\end{split}
\end{eqnarray}
namely no problem  when $\mu \neq 0$.

\section{Discussion}

In this paper, we have found an explicit covariant Lagrangian formulation of the loop quantum cosmology tree level  correction  \cite{Bojo} to Friedmann equation in terms of an infinite sum of non-polynomial gravity corrections to Einstein-Hilbert action. We have seen  that they constitute a suitable effective action in FLRW because, despite their non-polynomiality for general metric fields,  their contributions  evaluated on FLRW space-times are in fact polynomial. Furthermore, we have  found the exact solution of the model in presence of a positive  cosmological constant and a perfect fluid with state parameter $w$ and we have seen that it represents a bounce that replaces the big bang singularity of GR.

Our result  implements quite simply the argument of \cite{Helling_3,Date} consisting in interpreting  the LQC corrections as purely geometrical corrections to GR. In fact, the corrections lead to second order equations of motion in their FLRW sector, and therefore do not involve additional degrees of freedom with no direct geometrical interpretations, just like the Ricci and Gauss-Bonnet scalars that lead to second order equations of motion for general metric. In some sense, this might be expected since LQC is a quantum geometry theory, and this implies that, in a suitable limit,  it could be written as an effective action with initial term the Einstein-Hilbert one,
plus  high energy corrections with a direct geometrical interpretation.

For this effective action, such new degrees of freedom would be necessarily present since the Lovelock theorem \cite{Lovelock_1,Lovelock_2,Lovelock_3} prevents to find corrections, involving only the metric field, and with associated second order differential equations for any metric field (what excludes NPG). The effective equations of motion would therefore be higher order ones for general metric field. This  means that the theory would involve additional fields. If such an effective action formulation exists, it must therefore involve new degrees of freedom at least in some specific backgrounds.
In this sense, the NPG approach should be thought of as a convenient and simple way to explore  high energy corrections to the classical degrees of freedom only, in some specific backgrounds, in order to find effective (possibly non-singular) space-time solutions. The case of static spherically symmetric space-times, namely the case of regular black hole, is investigated  in \cite{RBHNPG}.

In our opinion, also the mimetic approach followed in \cite{Mukhanov,Mukhanov_2} and \cite{noui} might be an interesting step to understand a would-be modified gravity formulation of LQG semi-classical corrections, because, at least in a cosmological context, the additional field $\phi$ of this theory, due to the presence of a Lagrange multiplier, has a direct geometrical meaning $\phi=t$. This mimetic approach was also used  in \cite{mi_1,mi_2,mi_3,mi_4,mi_5,mi_6,mi_7}, and there might be a very large class of theories that have the same property to convert additional fields without a clear geometrical meaning into ones which are related to geometry via Lagrange multipliers.

In order to continue this work, one could try to generalize the construction for more general cosmological models like Bianchi I. In this special case, the construction of (\ref{Order0}) is preserved because once again $\partial_\alpha R$ has only one component, and therefore the NPG approach can be used in a simple way without searching first for new order-$0$ tensors.
However, in this way or even using mimetic gravity, one faces the issue raised earlier that an infinite number of scalars that do not contribute to FLRW dynamics (like those involving the Weyl tensor, boundary and vanishing terms that are such only in FLRW, etc) could be present. However, a straight and naive generalization of action (\ref{Action}) in the context of NPG could nontheless lead to interesting results.

We conclude by observing that it might be interesting, within this NPG approach, to see numerically if a bounce can be obtained at lowest order of corrections. With regard to the  action (\ref{Action}), we  used an infinite sum
$L_{NPG}^{\infty}= \sum_{i=0}^{\infty } \alpha_i \; S^i \; \Omega  $ and we had the necessity to reconstruct all the constants $\alpha_i$ given the result of LQC. If a bounce can already be obtained for a truncation at order $2(m+1)$ using the corrections $L_{NPG}^{m}= \sum_{i=0}^{m} \alpha_i \; S^i \; \Omega  $, it might  be possible to constraint the constant $\alpha_i$ for $i > 2(m+1)$ in order to preserve the bounce. This   could be an interesting way to single out a class of possible corrections without the need to  reconstruct the whole sequence of constants.

\acknowledgments{ This research has been supported by TIFPA-INFN  within INFN project FLAG.  We would like to thank the anonymous referee for  useful
 remarks. }

\end{document}